\documentstyle[preprint,aps]{revtex}
\begin{document}          
\title{Josephson Effect in BEC with Spin Degree of Freedom}
\author{S. Ashhab and Carlos Lobo}
\date{7/24/2001}
\maketitle
\begin{abstract} We consider the Josephson Effect between two
spatially separated Bose-Einstein condensates of atoms each of which can be in two
hyperfine states. We derive simple equations of motion for this system closely analogous
to the Bloch equations. We also map the dynamics of the system onto those of a classical
particle in a well. We find novel density and spin modes of oscillation and new stable
equilibrium points of the motion. Finally we analyze the oscillation modes in the spin-1
($F$=1) case.
\end{abstract}
\pagebreak
\section{Introduction} Recent experiments on the Josephson effect in $^3$He have
seen new phenomena that  have not been previously observed in conventional
superconducting junctions or in 
$^4$He superleaks \cite{Packard}. Some authors have attributed these
effects to the spin of the Cooper pairs since they are paired in an S=1 configuration 
\cite{Yip,Viljas}. By analogy this suggests that the simple Josephson effect in Bose
condensed alkali gases (BEC) could be qualitatively modified by the presence of
an internal spin degree of freedom. Recently several experimental groups have
succeeded in  forming this kind of condensate in which the atoms can be  in more
than one hyperfine state.  The JILA group \cite{JILA} has trapped the
$|F=2,m_F=1 \rangle$ and $|F=1,m_F=-1
\rangle$ states of $^{87}$Rb using magnetic fields. The Ketterle group at MIT
\cite{MIT} has trapped the $F=1$ multiplet of $^{23}$Na with optical methods. Both
groups have obtained condensates with densities around
$10^{14}$ cm$^{-3}$. An important feature of their setup is the possibility of
imaging each species separately, thus allowing them to observe their individual
motion. There have also been several theoretical studies of these systems
\cite{Ho,Ohmi,Bigelow,Koashi}.

The external Josephson effect, i.e. between two spatially separated condensates in the
case of a single hyperfine state, has already been addressed extensively in the
literature \cite{Giovanazzi,Smerzi,Milburn,Walls,Villain,Zapata,Cornell}. The
internal Josephson effect (between hyperfine states) has also been analyzed
\cite{Holland}. In the present work we study only the external Josephson effect, that
between two spatially separated condensates in a double well potential whose atoms have
two possible internal states, $|1 \rangle$ and $|2 \rangle$. A weak link is  established
between the condensates by lowering the potential barrier that  separates them.

This paper is organized as follows: first we describe the model and the
Hamiltonian of the system in Sec. II. After that, in Sec. III
we derive the  equations of motion and show that, in the so-called isotropic
limit, they reduce to  a form which is equivalent to the well-studied Bloch
equations. In Sec. IV we identify the equilibrium points of the motion and study their
dynamic stability. Finally, in Sec. V we make some considerations regarding
the extension  to the $F=1$ case (when the trapping potentials for all three $m_F$
sublevels are identical).

\section{MODEL SYSTEM AND HAMILTONIAN}

Let us take a condensate in a symmetric double well potential where the barrier is
much larger than the chemical potential and therefore, in order to go from one side
to the other, the atoms must tunnel under the barrier. Each atom has two
possible hyperfine states, which means that the order parameter is a two
component function. This setup can be achieved by taking a condensate in a
single well and raising a potential barrier in the middle, thereby splitting
it into two parts. Following this we may apply laser pulses to each
side selectively in order to choose a particular superposition of internal states of the
atoms on each side.

We shall now make a four mode approximation to describe the system.
Let $|1, R \rangle$, $|2, R\rangle$, $|1, L \rangle$ and $|2, L \rangle$ be the
four single atom states corresponding to the four modes, where the labels $R$ and $L$
refer to the right and left wells. The single atom
states are, in principle, time-dependent and will be approximately given, in the
adiabatic approximation, by the Gross-Pitaevskii groundstate, which in turn is determined
by the number of particles in each single atom state. When each of the four states
defined above is macroscopically occupied, the  condensate wavefunction $\Psi_{R,L}^i$
($i=1,2$) inside each of the wells can be well described at the Gross-Pitaevskii level.

We shall assume that the system is always in the semiclassical regime, i.e. that the
fluctuations around the mean values of the physical quantities are small (see below).
With this proviso we can describe the system in terms of classical (c-number)
canonically conjugate variables \cite{Tony}. The semiclassical variables are: $\Delta
N_i$ - one half of the difference in the number of atoms in each internal state, and
the relative phases $\Delta
\varphi_i$, which are canonically conjugate to each other. The condensate
wavefunction can be also expressed in terms of these variables as $\Psi_{R,L}^j=
\sqrt{N_j \pm \Delta N_j} \exp{(\pm i \Delta \varphi_j/2)}$ up to an overall phase, where
the $N_j$'s are one half of the total number of atoms in state $j$. Since we are
assuming in this paper that there is no laser coupling between the different spin
states, it follows that the different
$N_j$'s are conserved separately. Therefore the dynamics is not
naturally described by using as variables the relative phases between internal states in
each well.  Although the difference in chemical potential between the two species in
each well can be very large, this does not affect the Josephson
dynamics. For the same reason, also the dephasing between states of different hyperfine
spin is irrelevant.

Furthermore, if we restrict ourselves to values of $(\Delta N_1+\Delta
N_2)/(N_1+N_2)\ll 1$, we can write down an approximate Hamiltonian
 which is the straightforward generalization of
the spinless case and reduces to it when one of the $N_i$'s (and therefore the $\Delta
N_i$'s) goes to zero:

\begin{equation} H=H_{\rm J}+H_{\rm int} \label{eq:H}
\end{equation}

where

\begin{equation} H_{\rm J}=- \omega_0 \sum_{i=1,2}\sqrt{N_i^2-\Delta N_i^2} 
\cos \Delta \varphi_i 
\end{equation}

and

\begin{equation} H_{\rm int}=\frac{1}{2}\left( \epsilon_{11} \Delta
N_1^2+\epsilon_{22}
\Delta N_2^2+2 \epsilon_{12} \Delta N_1 \Delta N_2 \right)
\end{equation}

$H_{\rm J}$ is the Josephson coupling Hamiltonian and $H_{\rm int}$ is the interaction
Hamiltonian (for a derivation of these see \cite{Smerzi,Zapata}). The 
interaction term conserves the total number of atoms in each hyperfine state 
separately and we are here ignoring loss processes that occur in the  real system.
$\omega_0$ is the Josephson tunneling energy which we take to be the same for
both  hyperfine states for simplicity; and $\epsilon_{ij}$ are the effective
interaction coefficients. We assume that $\omega_0$ is independent of
$\Delta N_i$ and $\Delta \varphi_i$, although  we do expect some dependence 
for large values of $\Delta N_i$. We shall also assume that all $\epsilon$'s are
positive because we want to avoid two possible complications: the 
collapse of the gas and a possible accumulation of atoms on one side of the junction.

In the spinless case we can identify four different dynamical regimes as the parameters
$\omega_0$ and $\epsilon$ are varied. By analyzing the topology of the phase
space (Fig. 1a-b-c), we find two sharp transitions and one crossover. Starting from the
weakly interacting limit (the so-called Rabi regime - Fig. 1a), the transition to
the intermediate  regime is marked by the appearance of closed orbits oscillating around
nonzero values of
$\Delta N$ and centered at $\Delta \varphi=\pi$ (Fig. 1b). The second occurs when
open orbits appear where $\Delta \varphi$ extends over all values (Fig. 1c). This
is a transition from the intermediate to the Josephson regimes. These transitions happen
at the values $\omega_0=\epsilon N$ and $2 \omega_0= \epsilon N$ respectively for our
model. Finally when $\omega_0 N \sim \epsilon$ there is a crossover to the Fock regime
where quantum phase fluctuations cannot be neglected. Deep in the Rabi regime (when
$\omega_0 \gg \epsilon N$) the tunneling energy dominates. In the Josephson regime both
$H_{\rm J}$ and $H_{\rm int}$ are important ($\omega_0 \ll \epsilon N \ll
\omega_0 N^2$) and finally, in the Fock regime, $H_{\rm int}$ dominates ($\omega_0
N \ll \epsilon$) \cite{Tony}.

$\omega_0$ can be varied anywhere from 0 to 100 $s^{-1}$.
On the other hand, $\epsilon$ can go from 0.01 to 0.1 $s^{-1}$ \cite{Zapata} and,
under current experimental conditions, $N$ is usually between $10^4$ and 
$10^7$. With this range of parameters the Fock and Josephson regimes are easily
accessible whereas the Rabi regime is more difficult to achieve. It is important that
the frequency of any oscillation between the wells  be smaller than
the lowest intrawell excitation frequencies so that, during the motion, these degrees of
freedom are not excited. In practice this means that both $\omega_0$ and $\sqrt{\omega_0
\epsilon N}$ have to be smaller than the frequency of the lowest intrawell collective
mode (as will become clear below).

To be consistent with the semiclassical description we require that the
standard deviations of the quantum operators $\Delta \hat{\varphi}$ and $\Delta \hat{N}$
satisfy the conditions $\sigma( \Delta \hat{\varphi} )\ll 1$ and $\sigma(\Delta 
\hat{N}) \ll N$ during the motion of the system.  Generally speaking, the
experimental  setups will be such that $\sigma(\Delta \hat{N}) \sim \sigma^{-1}( \Delta
\hat{\varphi} ) \sim (\epsilon/\omega_0 N+1/N^2) ^{-1/4}$. The second
inequality is always satisfied for positive $\epsilon$'s. The first is satisfied
only in the Josephson and Rabi regimes, to which we shall restrict our analysis from now
on.

In order to justify the effective Hamiltonian the two components must be
miscible, in other words, there can be no component separation. If this were not
the case we might not have the same tunneling matrix element $\omega_0$
for both species and the mean field interaction energy would not have the
form that we assume. This means that a condition must be imposed on the
interaction parameters, namely that
$\epsilon_{11} \epsilon_{22}>\epsilon_{12}^2$ \cite{Ho2}.

Finally, the experimental observations can be made by measuring the density
(and therefore $\Delta N_{1,2}$) in the usual way, either destructively or by
phase-contrast imaging. As mentioned before, an important point is that these methods
allow us to determine experimentally the behaviour of each hyperfine species separately.

\section{DYNAMICS AND BLOCH EQUATIONS}

The equations of motion are:

\begin{eqnarray}
\dot{\Delta N_i} & = &- \frac{\partial H}{\partial \Delta \varphi_i}=- \omega_0
\sqrt{N_i^2-\Delta N_i^2}\sin \Delta \varphi_i \label{eq:motion1} \\ 
\dot{\Delta \varphi_i}& = & \frac{\partial H}{\partial \Delta N_i}=\omega_0
\frac{\Delta N_i}{\sqrt{N_i^2-\Delta N_i^2}} \cos \Delta
\varphi_i+ \sum_j \epsilon_{ij} \Delta N_j \label{eq:motion2}
\end{eqnarray}

In this section we will rewrite the equations of motion in terms of new variables in
order to provide some insight into the dynamics of the system.

\subsection{Isotropic case}

We first consider the isotropic case where $\epsilon_{11}= \epsilon_{12}=
\epsilon_{22}$. We shall be working with the quantity 

\begin{equation}
\epsilon \equiv \frac{1}{4}(\epsilon_{11}+\epsilon_{22}+2 \epsilon_{12})
\end{equation}

which is in fact equal to any of the $\epsilon$'s in the isotropic case. This
definition however, will be useful ahead, when we deal with the anisotropic situation.
The equality of the interaction parameters in the isotropic case seems to violate the
miscibility condition that $\epsilon_{11} \epsilon_{22}> \epsilon_{12}^2$. However this
condition does not take into account the kinetic energy which favours miscibility.
Therefore isotropy does not pose any such problems.

We notice that the Hamiltonian (\ref{eq:H}) is invariant under arbitrary spin rotations
and more generally of
$SU(2)$ transformations applied simultaneously to the spins in both wells. That is, if 
we transform the
two-component spinors $\Psi_L$ and $\Psi_R$ with the same unitary operator
then the dynamics should remain unchanged. This suggests that we re-express
the equations of motion in terms of quantities which are invariant under such
transformations. This conclusion of course depends on the isotropy of
the interaction Hamiltonian
$H_{\rm int}$. We therefore define the following dot products of spinors:

\begin{eqnarray}
\Delta N_+ & \equiv & \frac{|\Psi_R|^2-|\Psi_L|^2}{2} = \sum_i \Delta N_i\\
\alpha_+ & \equiv & \frac{\Psi_L^* \Psi_R- \Psi_R^*\Psi_L}{2i} = \sum_i 
\sqrt{N_i^2-\Delta N_i^2} \sin \Delta \varphi_i\\
\beta_+ & \equiv & \frac{\Psi_L^* \Psi_R+ \Psi_R^*\Psi_L}{2} = \sum_i
\sqrt{N_i^2-\Delta N_i^2} \cos \Delta \varphi_i
\end{eqnarray}

The subscript (+) will be used to distinguish this set of variables from another one with
subscript ($-$) to be defined below. Using the equations of motion for 
$\Delta N_i$ and $\Delta \varphi_i$, we obtain

\begin{equation}
\pmatrix{\dot{\Delta N_+}\cr \dot{\alpha_+}\cr \dot{\beta_+}}=  
\label{eq:bloch1}
\pmatrix{0 & -\omega_0 & 0 \cr 
         \omega_0 & 0 & +\epsilon \Delta N_+ \cr
         0 & -\epsilon \Delta N_+ & 0 \cr}
\pmatrix{\Delta N_+\cr \alpha_+\cr \beta_+} \label{eq:Bloch1}
\end{equation} 

If we now define the three-component vectors
${\bf r_+}=(\Delta N_+,\alpha_+,\beta_+)$ and ${\bf B}(t)=(- \epsilon \Delta
N_+,0,\omega_0)$ then we can rewrite the equations of motion succinctly as

\begin{equation}
{\bf \dot{r}_+=B}(t) {\bf\times r_+} \label{eq:Bloch1a}
\end{equation}

Note though that ${\bf{B}}$ and ${\bf{r_+}}$ are not independent
since they are both functions of $\Delta N_+$.

Straightforward manipulation of Eq. (\ref{eq:Bloch1}), or directly of
Eqs. (\ref{eq:motion1}) and (\ref{eq:motion2}), leads to

\begin{equation}
\ddot{\Delta N_+}=-\left(\omega_0^2 - \epsilon H(0)
 \right) \Delta N_+ - \frac{\epsilon^2}{2}
\Delta N_+^3
\end{equation}

where $H(0)= -\omega_0 \beta_+(0)+\frac{\epsilon}{2} \Delta N_+(0)^2$, ($N=\sum_i N_i$)

This equation is quite general since it is valid not only for two hyperfine
states but for any number of them, as long as they interact only through a
$\Delta N_+^2 \equiv (\sum_i
\Delta N_i)^2$ term. In particular it also applies to a single state (i.e. spinless)
system. It is formally identical to the equation of motion of a particle 
with unit mass in the quadratic-plus-quartic effective potential

\begin{equation} 
V_{eff}(\Delta N_+)=\frac{1}{2} \left(\omega_0^2 - \epsilon H(0) \right) \Delta N_+^2+
\frac{\epsilon^2}{8} \Delta N_+^4 \label{eq:Veff}
\end{equation}

with effective total energy 

\begin{equation} E_{eff} = V_{eff}(\Delta N_+) + \frac{1}{2} \dot{\Delta
N_+}^2
\end{equation}

An important point to notice is that $E_{eff}$ and $V_{eff}$ cannot be chosen
independently since they both depend on the initial conditions. The variation of $H(0)$
and of $\alpha_+(0)$ (since $\dot{\Delta N}=-\omega_0 \alpha_+$) allows us to find three 
different types of motion (Fig. 2a-b-c). In the first type the coefficient of the
quadratic term is positive and
$\Delta N_+$ oscillates around zero, which is the minimum of $V_{eff}$ (Fig. 2a). In the
spinless case this corresponds to either oscillations around the
origin (Fig. 1a-b-c) or to small oscillations around the $\pi$-state (Fig. 1a). The
second case occurs when the coefficient is negative and $E_{eff}$ is positive, which
also leads to oscillations of $\Delta N_+$ around zero although that point is no longer a
minimum of $V_{eff}$ (Fig. 2b). It corresponds to large oscillations around
the origin or the $\pi$-state (Fig. 1b-c). The third one corresponds to both $E_{eff}$
and the coefficient being negative (Fig. 2c) and leads to self-trapped behaviour
(oscillations around $\Delta N_+ \neq 0$ - Fig. 1b-c). For the spinless case there is a
well-known analogy with a momentum-shortened pendulum in a gravitational field whose
behaviour is also fully reproduced by this particle-in-a-well model. Since the  analysis
of the spinless junction has already been carried out in Ref. \cite{Smerzi} we shall not
continue it here and shall proceed to the two hyperfine state case.

Specifying the dynamics of $\Delta N_+$ does not describe the motion
completely. For example, even in the spinless case it is known that the third regime
includes two different behaviours of the relative phases, the so-called ``running" and
``oscillating" phases. For a description of these as well as $\pi$-states and the
momentum-shortened pendulum analogy see e.g. Ref. \cite{Smerzi}. To further understand
the dynamics of the two hyperfine state Josephson effect we introduce the
additional variables

\begin{eqnarray}   
\Delta N_- & \equiv & \Delta N_1 - \Delta N_2\\  
\alpha_- & \equiv & \sqrt{N_1^2-\Delta N_1^2} \sin \Delta \varphi_1-
\sqrt{N_2^2-\Delta N_2^2} \sin \Delta \varphi_2\\
\beta_- & \equiv &  \sqrt{N_1^2-\Delta N_1^2} \cos \Delta \varphi_1-
\sqrt{N_2^2-\Delta N_2^2} \cos \Delta \varphi_2
\end{eqnarray}

and their equations of motion are:

\begin{equation}
\pmatrix{\dot{\Delta N_-}\cr \dot{\alpha_-}\cr
\dot{\beta_-}}=
\pmatrix{0 & -\omega_0 & 0 \cr 
         \omega_0 & 0 & +\epsilon \Delta N_+ \cr
         0 & -\epsilon \Delta N_+ & 0 \cr}
\pmatrix{\Delta N_-\cr \alpha_-\cr \beta_-}
\end{equation} 

Since the matrix is the same as in Eq. (\ref{eq:Bloch1}) we define the
three-component vector ${\bf r_-}=(\Delta N_-,\alpha_-,\beta_-)$ and rewrite
the equations of motion as

\begin{equation} {\bf \dot{r}_-=B}(t) {\bf\times r_-} \label{eq:Bloch2}
\end{equation}

Now, however, ${\bf B}$ and ${\bf r_-}$ are independent and therefore these
equations are formally identical to the Bloch equations, familiar from the
context of NMR and quantum optics.
Notice that in going from the original four variables to six we are enlarging 
the configuration space, which means that not all points described by the new
set of variables are physically allowed. Therefore care must taken in choosing
the initial conditions of the motion.

We can obtain some physical insight into the variables $\Delta N_{\pm}$ by noting that
$\Delta N_+$ is one half of the difference in total number between the right and left
wells and, in the limit $N_1=N_2$, $\Delta N_-$ is one half of the difference in spin
between the wells
$(|\Psi^1_R|^2-|\Psi^2_R|^2)/2-(|\Psi^1_L|^2-|\Psi^2_L|^2)/2$. This means that the former
describes the density mode whereas the latter, in that limit, describes the spin mode. 

We can now analyze Eqs. (\ref{eq:Bloch1a}) and (\ref{eq:Bloch2}) in a few limiting cases
to gain some insight into the behaviour of the system. It is possible to have small
oscillations in
$\Delta N_+$ and no motion in $\Delta N_-$ or vice-versa. We consider two cases: the
Rabi limit where $\epsilon N \ll \omega_0$ and the Josephson regime, where $\epsilon N
\gg
\omega_0$. In the Rabi case, neglecting higher order terms, the frequency of
oscillation of
$\Delta N_+$ (and therefore of the density mode) can be calculated from
Eq. (\ref{eq:Veff}) to be $\sqrt{\omega_0^2+\omega_0
\epsilon \beta_+}$. Also, using Eq. (\ref{eq:Bloch2}), we can neglect the component of
the
${\bf B}$ field along the $\Delta N_-$ axis so that ${\bf r_-}$ (the spin mode) rotates
around the $\beta_-$ axis with frequency $\omega_0$. In the Josephson case we
consider two types of situations - small oscillations of
$\Delta N_+$ around zero and around nonzero values. For zero values we can get 
density and spin modes with frequencies $\sqrt{\omega_0^2+\omega_0 \epsilon N}$ and
$\omega_0$ respectively. For nonzero values (i.e. when $\Delta N_+$ is 
``self-trapped'' around a value $\Delta N_+^0$), $\Delta N_+$ oscillates with frequency
$\epsilon |\Delta N_+^0|$ and $\Delta N_-$ with frequency $\sqrt{\omega_0^2+\epsilon^2
(\Delta N_+^0)^2}$.

\subsection{Anisotropic case}

As we would expect, the equations in this case become much more complicated.
However, for the sake of completeness, we include them here. We find
that the equations for ${\bf \dot{r}_+}$ and ${\bf \dot{r}_-}$ become coupled:

\begin{eqnarray} {\bf \dot{r}_+=B_1}(t) {\bf\times r_+} +{\bf B_2}(t)
{\bf\times r_-}\\ {\bf \dot{r}_-=B_1}(t) {\bf\times r_-}+{\bf B_2}(t)
{\bf\times r_+}
\end{eqnarray}

where ${\bf B_1}=(-\epsilon \Delta N_+-\epsilon_B \Delta N_-,0,\omega_0)$,
${\bf B_2}=(-\epsilon_A \Delta N_--\epsilon_B \Delta N_+,0,0)$,
$\epsilon \equiv \frac{1}{4}(\epsilon_{11}+\epsilon_{22}+2\epsilon_{12}$) as
before,
$\epsilon_A \equiv \frac{1}{4}(\epsilon_{11}+\epsilon_{22}-2\epsilon_{12})$
and 
$\epsilon_B \equiv \frac{1}{4}(\epsilon_{11}-\epsilon_{22})$.

\section{DISCUSSION OF THE EQUILIBRIUM POINTS OF THE MOTION}

In this section we study the existence and stability of the equilibrium points of the
motion. The main results for the isotropic case are summarized in Table 1.

At an equilibrium point, $\dot{\Delta N_{1,2}}=\dot{\Delta \varphi_{1,2}}=0$.
Using Eq. (\ref{eq:motion1}) this implies that the phases $\Delta \varphi_1$ and
$\Delta \varphi_2$ are either zero or $\pi$. From Eq. (\ref{eq:motion2}) we get

\begin{eqnarray}
\Delta N_1^0=-\Delta N_2^0 \left(\frac{\epsilon_{22}}{\epsilon_{12}}+
\frac{\omega_0}{\epsilon_{12} \sqrt{N_2^2-(\Delta N_2^0)^2} \zeta_2} \right) 
\label{eq:equilibrium1}\\
\Delta N_2^0=-\Delta N_1^0 \left(\frac{\epsilon_{11}}{\epsilon_{12}}+
\frac{\omega_0}{\epsilon_{12} \sqrt{N_1^2-(\Delta N_1^0)^2} \zeta_1} \right) 
\label{eq:equilibrium2}
\end{eqnarray}

where $\Delta N_i^0$ and $\Delta \varphi_i^0$ are the coordinates of the equilibrium
point. We have defined $\zeta_1 \equiv \cos \Delta \varphi_1^0$ and 
$\zeta_2 \equiv \cos \Delta \varphi_2^0$ to abbreviate the formulae.

These equations define two functions, $\Delta N_1^0(\Delta N_2^0)$ and 
$\Delta N_2^0(\Delta N_1^0)$, which we can plot on the ($\Delta N_1^0$, 
$\Delta N_2^0$) plane (Fig.3).
Since both $\zeta_1$ and $\zeta_2$ can be $1$ or $-1$ (corresponding to 
$\Delta \varphi^0=0$ or $\pi$) we have three distinct cases. For all of them the trivial
point $\Delta N_1^0=\Delta N_2^0=0$ is always a solution. Therefore in all cases we have 
at least one solution.

Case 1: $\zeta_1=\zeta_2=1$ ($\Delta \varphi_1=\Delta \varphi_2=0$)

As is clear from Fig. 3a-b the condition for the existence of three solutions 
imposes a condition on the slopes of the curves at the origin which leads to 

\begin{equation}
\left(\frac{\epsilon_{22}}{\epsilon_{12}}+ \frac{\omega_0}{\epsilon_{12} N_2} 
\right) \left(\frac{\epsilon_{11}}{\epsilon_{12}} +
\frac{\omega_0}{\epsilon_{12} N_1} \right)< 1 \label{eq:case1}
\end{equation}

Case 2: $\zeta_1=-1$, $\zeta_2=1$ ($\Delta \varphi_1=\pi$, $\Delta 
\varphi_2=0$)

For three solutions to exist (Fig. 3c-d) we require this time that

\begin{equation}
\left(\frac{\epsilon_{11}}{\epsilon_{12}}- \frac{\omega_0}{\epsilon_{12} N_1}
\right)
 \left(\frac{\epsilon_{22}}{\epsilon_{12}} +
\frac{\omega_0}{\epsilon_{12} N_2} \right) >  1 \label{eq:case2}
\end{equation}

Case 3: $\zeta_1=\zeta_2=-1$ ($\Delta \varphi_1=\Delta \varphi_2=\pi$)

This time we may have one (Fig. 3e), three (Fig. 3f) or five (Fig. 3g) solutions. To have
three we need that

\begin{equation}
\left(\frac{\epsilon_{22}}{\epsilon_{12}}- \frac{\omega_0}{\epsilon_{12} N_2} 
\right) \left(\frac{\epsilon_{11}}{\epsilon_{12}} -
\frac{\omega_0}{\epsilon_{12} N_1} \right)< 1 \label{eq:case3}
\end{equation}
If this condition is not met then we will have one or five solutions, 
depending on whether the factors on the left hand side are both negative or both
positive respectively.

In the rest of this section we will analyze the behaviour of the system
close  to the various equilibrium points. Notice that the global groundstate is the
trivial solution $\Delta N_{1,2}=0$ and $\Delta \varphi_{1,2}=0$. All the other
equilibrium points are thermodynamically unstable although possibly  dynamically stable
\cite{Castin}.

\subsection{Isotropic case}

If all $\epsilon_{ij}$'s are equal then some of the conditions in the previous
subsection cannot be satisfied. For Case 1, condition (\ref{eq:case1}) cannot
be satisfied and therefore only the  equilibrium point $\Delta N_1=\Delta
N_2=0$ is allowed.

In Case 2 both single and triple solutions are allowed. Condition
(\ref{eq:case2}) for the existence of three equilibrium points becomes
$N_1-N_2-\omega_0/\epsilon>0$.

In Case 3 the conditions for the existence of one or three equilibrium points
can be satisfied. The condition for three is $N_1+N_2-\omega_0/\epsilon>0$. However we
cannot have five equilibrium points since, if both terms are positive in condition
(\ref{eq:case3}) and each of them is smaller than one, their product will
also be smaller than one.

To study the behaviour in the neighbourhood of an equilibrium point we shall
work with  the second order differential equations for $\Delta N_1$ and $\Delta N_2$. To
obtain these we differentiate
Eq. (\ref{eq:motion1}) with respect to time and eliminate $\dot{\Delta
\varphi_{1,2}}$ and $\dot{\Delta N_{1,2}}$ using Eqs.
(\ref{eq:motion1},\ref{eq:motion2}). We now introduce the variables 
$\delta_1,\delta_2$ defined by

\begin{eqnarray}
\Delta N_1 = \Delta N_1^0 +\delta_1 \\
\Delta N_2 = \Delta N_2^0 + \delta_2 
\end{eqnarray}

The linearized equations of motion for the isotropic case are:

\begin{equation}
\pmatrix{\ddot{\delta_1} \cr \ddot{\delta_2} \cr}=-{\bf \Omega}^2
\pmatrix{\delta_1 
\cr \delta_2 \cr} \label{eq:delta}
\end{equation}

where

\begin{equation} {\bf \Omega}^2 = \left( \omega_0^2+\epsilon^2 (\Delta N_+^0)^2 \right)
{\bf Id}+  \omega_0 \epsilon
\pmatrix{\sqrt{N_1^2-(\Delta N_1^0)^2} \zeta_1 & 
\sqrt{N_1^2-(\Delta N_1^0)^2} \zeta_1 \cr
\sqrt{N_2^2-(\Delta N_2^0)^2} \zeta_2 &
\sqrt{N_2^2-(\Delta N_2^0)^2} \zeta_2 \cr}
\end{equation}
\

Case 1: $\zeta_1=\zeta_2=1$

As mentioned before, the only stable point is at $\Delta N_1^0=\Delta
N_2^0=0$. In the basis
$(\delta_1,\delta_2)$ we find the modes $(N_1,N_2)$ and $(1,-1)$ (note that
the matrix of the linearized equations of motion is not hermitian and
therefore the two eigenvectors are not guaranteed to be orthogonal even if
the corresponding frequencies are different). The first corresponds to a 
`density' mode with frequency $\sqrt{\omega_0^2+\omega_0
\epsilon N}$ and the second to a `spin' mode with frequency $\omega_0$. Note though
that, even in the density mode, the total spin on each side of the junction changes as a
function of time (unless $N_1=N_2$).

\

Case 2: $\zeta_1=-1$, $\zeta_2=1$

Near $\Delta N_1^0=\Delta N_2^0=0$ we proceed as above and find the
eigenfrequencies
$\sqrt{\omega_0^2+\omega_0 \epsilon (N_2-N_1)}$ and $\omega_0$ with
corresponding eigenvectors $(N_1,-N_2)$ and $(1,-1)$.
The system is dynamically stable as long as the frequencies are real, which leads
to the condition $N_1-N_2-\omega_0/\epsilon<0$. Since $N_1$ and $N_2$ are easy to
change experimentally, this state can always be made stable regardless of the
values of $\omega_0$ and $\epsilon$. It is therefore much easier to obtain a $\pi$
state this way than in the spinless case.  However there is an additional
complication: if
$N_1=N_2$ the eigenvectors become parallel and, since the representation of
arbitrary vectors in an almost colinear basis can involve very large
amplitudes (especially for the vectors perpendicular to the basis vectors),
the amplitude of the oscillations for initial displacements in the in-phase
direction will tend to diverge as $(N_2 - N_1)/N \rightarrow 0$ after some time.
Experimentally it is not difficult to avoid this pitfall. If $\Delta N_1^0
\neq 0$  and $\Delta N_2^0\neq 0$ the frequencies become
$\epsilon |\Delta N_+^0|$ for the density mode and 
$\sqrt{\omega_0^2+\epsilon^2 (\Delta N_+^0)^2}$ for the spin
mode. Both frequencies are real and therefore this equilibrium point is stable.
\

Case 3: $\zeta_1=\zeta_2=-1$

For $\Delta N_1^0=\Delta N_2^0=0$ we find the eigenfrequencies
$\sqrt{\omega_0^2-\omega_0 \epsilon N}$ and $\omega_0$ with corresponding
eigenvectors $(N_1,N_2)$ and $(1,-1)$. The system is dynamically stable as 
long as $N<\omega_0/\epsilon$. If $\Delta N_1^0,\Delta N_2^0 \neq 0$ then the
frequencies of the two eigenmodes are given by the same expressions as those for $\Delta
N_1^0,\Delta N_2^0 \neq 0$ in Case 2 ($\epsilon |\Delta N_+^0|$ and 
$\sqrt{\omega_0^2+\epsilon^2 (\Delta N_+^0)^2}$) and therefore they are both stable
as long as the points exist.

Note that all the frequencies found in Sec. III A are in agreement with those derived
here by studying the small oscillation behaviour directly from Eqs.
(\ref{eq:motion1},\ref{eq:motion2}).

\subsection{Anisotropic case}

In the anisotropic case we can still use Eq. (\ref{eq:delta}) but with
${\bf\Omega}^2$  given by

\begin{eqnarray} 
{\bf \Omega}^2  & = & 
\pmatrix{\omega_0^2+ (\epsilon_{11} \Delta N_1^0 +\epsilon_{12} \Delta N_2^0)^2 & 0 \cr
0 & \omega_0^2+ (\epsilon_{22} \Delta N_2^0 +\epsilon_{12} \Delta N_1^0)^2}
\nonumber \\ \nonumber \\
& + & \omega_0
\pmatrix{\epsilon_{11} \sqrt{N_1^2-(\Delta N_1^0)^2} \zeta_1 &
\epsilon_{12}
\sqrt{N_1^2-(\Delta N_1^0)^2} \zeta_1 \cr
\epsilon_{12} \sqrt{N_2^2-(\Delta N_2^0)^2} \zeta_2 & \epsilon_{22} 
\sqrt{N_2^2-(\Delta N_2^0)^2} \zeta_2 \cr}
\end{eqnarray}
\

Case 1: $\zeta_1=\zeta_2=1$

When $\Delta N_1^0=\Delta N_2^0=0$ the eigenvalues are

\begin{equation}
\omega^2=\omega_0^2+\omega_0 \left( \frac{\epsilon_{11} N_1 \zeta_1+
\epsilon_{22} N_2 \zeta_2}{2} \pm \sqrt{\frac{(\epsilon_{11} N_1
\zeta_1-\epsilon_{22} N_2 \zeta_2)^2}{4} + \epsilon_{12}^2 N_1 N_2
\zeta_1 \zeta_2} \right) \label{eq:frequencies}
\end{equation}

with $\zeta_1=\zeta_2=1$.
For simplicity we shall address only the nearly isotropic case. It is
experimentally relevant since this is the case for $^{23}$Na where
the experimental values for the
$\epsilon_{11},\epsilon_{22}$ and
$\epsilon_{12}$ are similar. To do this we use the variables $\epsilon,\epsilon_{A,B}$
defined in section III B since
$\epsilon_A$ and $\epsilon_B$ quantify the degree of anisotropy. We
therefore  treat them as small parameters. Expanding the square root in
Eq. (\ref{eq:frequencies}) and keeping terms to first order in those variables we
obtain the two eigenvalues

\begin{equation}
\omega^2=\omega_0^2+\omega_0 \epsilon(N_1\zeta_1+N_2\zeta_2)+\omega_0 \epsilon_A
\frac{(N_1\zeta_1-N_2\zeta_2)^2}{N_1\zeta_1+N_2\zeta_2}+2 \omega_0
\epsilon_B (N_1\zeta_1 -N_2\zeta_2) \label{eq:asym2}
\end{equation}

\begin{equation}
\omega^2=\omega_0^2+4 \omega_0 \epsilon_A\frac{ N_1 N_2 \zeta_1 \zeta_2}{N_1\zeta_1+
N_2\zeta_2} \label{eq:asym1}
\end{equation}

We have assumed that $N_1+N_2\sim N_1-N_2 \sim N$.
Since $\epsilon_A>0$ (which is implied by the miscibility condition) both
modes  are stable ($\epsilon_B \ll \epsilon$). The instability that would arise
at  sufficiently large and negative values of $\epsilon_A$ has the same
origin as the immiscibility  condition. However we do not consider this region in
this paper since immiscibility would have severe consequences (see end of section II).
It is easy to see from Eq. (\ref{eq:case1}) that the case $\Delta N_1^0,\Delta N_2^0
\neq 0$ is also ruled  out due to the miscibility
condition.

\
 
Case 2: $\zeta_1=-1,\zeta_2=1$

At the origin $\Delta N_1^0=\Delta N_2^0=0$ the frequencies are again given 
by Eqs. (\ref{eq:asym2}) and (\ref{eq:asym1}) but with
$\zeta_1=-1,\zeta_2=1$. Let us divide the region into two parts: $N_1-N_2>0$ and
$N_1-N_2<0$. In the first region the motion is always unstable: for large values of
$N_1-N_2$ the first frequency is imaginary and, for small values, the resonance of
section IV A will  tend to destabilize the equilibrium point.
In the second region, if $N_1-N_2>-4 \epsilon_A N_1 N_2/\omega_0$ then again
it is unstable. Otherwise it is stable (provided it is outside the region of
resonance). For $\Delta N_1^0,\Delta N_2^0 \neq 0$,
the  conditions for stability become rather complex and offer little insight.
However, on physical grounds,  by suitably choosing the
parameters, any of the equilibrium points can be made stable.

\

Case 3: $\zeta_1=\zeta_2=-1$

The frequencies at the point $\Delta N_1^0=\Delta N_2^0=0$ are those given by 
Eqs. (\ref{eq:asym2}) and (\ref{eq:asym1}) with $\zeta_1=-1,\zeta_2=-1$.
As in the isotropic case, the point $\Delta N_1^0=\Delta N_2^0=0$ is
unstable for typical experimental conditions (see below), namely, when $\epsilon
N>\omega_0$. The corrections to this criterion are of order
$(\epsilon_A,\epsilon_B)/\epsilon$.
Finally, when $\Delta N_1^0,\Delta N_2^0 \neq 0$, as in the preceding case, the
stability can generally be achieved for all equilibrium points for appropriate values of
the parameters barring immiscibility problems.

\subsection{Experimental considerations}

The typical frequencies of small
oscillations can be calculated using the following parameters: $N \sim 10^6$
atoms, $\epsilon \sim 0.01$ s$^{-1}$,
$\epsilon_{A,B} \sim 10^{-4}$ s$^{-1}$ and $\omega_0 \sim 10$ s$^{-1}$. For 
these values most of the frequencies lie between 10 s$^{-1}$ and $100$ 
s$^{-1}$, whenever stable oscillations exist.

For a general initial state near the trivial equilibrium point, which is $\Delta
\varphi_{1,2}=0$ and $\Delta N_{1,2}=0$, the oscillations in $\Delta N_{1,2}$ and $\Delta
\varphi_{1,2}$ will be a superposition of both density and spin modes. For the typical
parameters that we are using, the frequency of the density mode is one order of magnitude
larger than that of the spin mode and, therefore, it should be simple to distinguish
between them experimentally. It should also be possible to prepare an initial state in
which only one of the two modes is significantly excited.

Although the $\pi$-states are unstable in the spinless case for typical parameters, we
have shown that they can be stabilized in spinor condensates. To prepare them
experimentally we must have $N_1<N_2$ as explained above, where the $\pi$-phase
difference is in species 1. A frequency measurement of the density mode could be used to
detect that in fact a $\pi$ phase exists in species 1. Alternatively, one could observe
the destabilization of the state suddenly appearing in the form of density
oscillations due to the reduction of
$N_2$. Finally, a third possibility would be the direct imaging of the interference
pattern between the left and right condensates of species 1 during its expansion, after
the trapping potentials have been switched off.

\section{$F$=1 SPIN JOSEPHSON EFFECT}

In this section we look at the mean-field groundstate and Hamiltonian of a Josephson
junction containing atoms with $F=1$ total spin. The groundstate of a 
single spinor condensate has been analyzed in the literature
\cite{Ho,Ohmi,Bigelow,Koashi}. Here we extend the analysis to the case where
the condensate is comprised of two spatially separated parts linked by a
weak junction. Some of the results in this section are similar to those of Ref.
\cite{Ho} and are related to the bulk excitation spectrum of spin-1 condensates. As in
the previous sections we assume that the trapping potentials are identical for all three
hyperfine states (which can be achieved using optical dipole traps). Under these
conditions it is known that all three hyperfine states of the multiplet are miscible.

One might try to proceed as in section III by deriving a set of equations for invariant
quantities such as $\Delta N_+$, $\alpha_+$, $\beta_+$ and so on. However it turns out
that while this is possible, it does not lead to simple equations of motion as
in the two internal state system and therefore this approach does not seem to provide a
clear insight into the dynamics.

We shall now study the small oscillations around some of the equilibrium
points in  both the ferro- and anti-ferromagnetic cases.

\begin{equation}
H=H_{\rm J} + H_{\rm int}
\end{equation}
 
where

\begin{equation} 
H_{\rm J}=-\frac{\omega_0}{2} \sum_i a^{\dagger}_{i,L} a_{i,R} + h.c.
\end{equation}

and \cite{Bigelow}

\begin{eqnarray} H_{\rm int}=\sum_{i=R,L} &&\frac{\epsilon_0}{4} ( 
a^{\dagger}_{1,i} a^{\dagger}_{1,i} a_{1,i} a_{1,i} + a^{\dagger}_{0,i}
a^{\dagger}_{0,i} a_{0,i} a_{0,i}+ a^{\dagger}_{-1,i} a^{\dagger}_{-1,i}
a_{-1,i} a_{-1,i} \nonumber \\ &&+ 2a^{\dagger}_{1,i} a^{\dagger}_{0,i}
a_{0,i} a_{1,i}+ 2a^{\dagger}_{0,i} a^{\dagger}_{-1,i} a_{-1,i} a_{0,i}+
2a^{\dagger}_{1,i} a^{\dagger}_{-1,i} a_{-1,i} a_{1,i} ) \nonumber \\
&&+\frac{\epsilon_2}{4} (  a^{\dagger}_{1,i} a^{\dagger}_{1,i} a_{1,i} a_{1,i}
+ a^{\dagger}_{-1,i} a^{\dagger}_{-1,i} a_{-1,i} a_{-1,i}+ 2a^{\dagger}_{1,i}
a^{\dagger}_{0,i} a_{0,i} a_{1,i} \nonumber \\ &&+ 2a^{\dagger}_{0,i}
a^{\dagger}_{-1,i} a_{-1,i} a_{0,i}+ -2a^{\dagger}_{1,i} a^{\dagger}_{-1,i}
a_{-1,i} a_{1,i}+ 2a^{\dagger}_{1,i} a^{\dagger}_{-1,i} a_{0,i} a_{0,i}
\nonumber \\ &&+ 2a^{\dagger}_{0,i} a^{\dagger}_{0,i} a_{1,i} a_{-1,i}-
a^{\dagger}_{1,i} a_{1,i}-a^{\dagger}_{-1,i} a_{-1,i} )
\end{eqnarray}

For e.g. $^{23}$Na, $\epsilon_2$ is approximately 2\% of $\epsilon_0$.

We now derive the equations of motion and, in the mean-field approximation, since we are
assuming a macroscopic occupation, we linearize by keeping only terms at least of order
$N$ in $H_{{\rm int}}$.

For the ferromagnetic case, if we assume that only the $m_F=1$ has macroscopic
occupation, we obtain the equations:

\begin{equation} i \frac{d}{dt} \pmatrix{\delta \phi_{1,L} \cr \delta
\phi_{0,L} \cr 
\delta \phi_{-1,L}}=-\frac{\omega_0}{2}
\pmatrix{\delta \phi_{1,R} \cr \delta \phi_{0,R} \cr \delta \phi_{-1,R}}+ 
\frac{N}{2}
\pmatrix{\epsilon_0+\epsilon_2 & 0 & 0 \cr 0 & \epsilon_0+\epsilon_2 & 0
\cr 0 & 0 & \epsilon_0-\epsilon_2} \pmatrix{2\delta \phi_{1,L} +  
\delta \phi_{1,L}^*\cr 
\delta \phi_{0,L} \cr \delta \phi_{-1,L}}
\end{equation}

and a similar set for $\delta \phi_{i,R}$.

Solving them gives the following results: for the groundstate (the relative 
phase between $\phi_{1,R}$ and
$\phi_{1,L}$ equal to zero), we have the following three modes: a density
mode with frequency $\sqrt{\omega_0^2+\omega_0 (\epsilon_0+\epsilon_2) N}$, a
spin mode with frequency $\omega_0$ and a quadrupole mode with frequency
$\omega_0+|\epsilon_2|N$.

For the $\pi$-state we find the same modes with frequencies 
$\sqrt{\omega_0^2-\omega_0 (\epsilon_0+\epsilon_2) N}$, $-\omega_0$ and
$-\omega_0+|\epsilon_2|N$. The density mode can clearly become unstable for
$\omega_0<(\epsilon_0+\epsilon_2) N$, which is the case for the typical
parameters quoted in the previous section.

For the anti-ferromagnetic case, if we assume that only the $m_F=0$ state is
macroscopically occupied, we obtain the equations:

\begin{equation} 
i \frac{d}{dt} \pmatrix{\delta \phi_{1,L} \cr \delta
\phi_{0,L} \cr 
\delta \phi_{-1,L}}=-\frac{\omega_0}{2}
\pmatrix{\delta \phi_{1,R} \cr \delta \phi_{0,R} \cr \delta \phi_{-1,R}}+
\frac{N}{2}
\pmatrix{\epsilon_0+\epsilon_2 & 0 & 0 \cr 0 & \epsilon_0 & 0
\cr 0 & 0 & \epsilon_0 + \epsilon_2} \pmatrix{\delta \phi_{1,L}   
\cr 2 \delta \phi_{0,L}+\delta \phi_{0,L}^* \cr \delta \phi_{-1,L}}
+\frac{N \epsilon_2}{2} \pmatrix{\delta \phi_{-1,L}^*   
\cr 0 \cr \delta \phi_{1,L}^*}
\end{equation}

and a similar set for $\delta \phi_{i,R}$.

Solving them gives the following results: for the groundstate (the relative 
phase between $\phi_{0,R}$ and
$\phi_{0,L}$ equal to zero), we have the following three modes: a density
mode with frequency $\sqrt{\omega_0^2+\omega_0 \epsilon_0 N}$ and two 
degenerate spin modes with frequency $\sqrt{\omega_0^2+\omega_0
\epsilon_2 N}$.

For the $\pi$-state we find the same modes with frequencies 
$\sqrt{\omega_0^2-\omega_0 \epsilon_0 N}$ and $\sqrt{\omega_0^2-\omega_0
\epsilon_2 N}$. The density mode becomes unstable for
$\omega_0<\epsilon_0 N$ and the spin modes become unstable for 
$\omega_0<\epsilon_2 N$. With the parameters that we are
using at the very least the density mode is unstable.

\section{Conclusions}

In this paper we considered the Josephson junction between two spatially
separated condensates with a hyperfine degree of freedom.  We have derived a
set of simple equations which, in the isotropic limit, are formally identical to the
Bloch equations and which provide insight into the dynamics of the two hyperfine state
condensate in a  double-well setup. We find a partial mapping to the simple
problem of a particle in a $\pm x^2 +x^4$ type potential which becomes a complete
mapping in the spinless case. We have also demonstrated the existence in this system of
new density and spin oscillation modes. In particular we have found
$\pi$-states that are  stable under experimentally accessible conditions due to the
interactions between the two species. Finally we analyzed the spin-1 case in the same
geometry both for the ferro- and anti-ferromagnetic cases and found the low-lying
oscillation modes. Our results indicate a wide range of  phenomenology for Josephson
oscillations when the superfluid has a spin degree of freedom. Future possible
directions of research might include tunneling between fragmented  states
\cite{Oktel} and more general solutions of the Bloch equations.

We would like to thank A. J. Leggett for many insightful comments and especially for
drawing our attention to this problem. We would also like to thank N. Hatakenaka and I.
Zapata for several stimulating discussions. Additionally we received useful advice from
G. Baym, Y. Castin, S. Kurihara, E. Mueller and G. S. Paraoanu. This work was supported
by the NSF grant no. DMR99-86199. CL acknowledges a PRAXIS XXI fellowship. SA would like
to thank N. Hatakenaka and H. Takayanagi for their hospitality during his stay at NTT
Basic Research Laboratories and acknowledges support from NTT Corporation.

\pagebreak

\begin{center}
FIGURES
\end{center}

FIG. 1. Orbits in phase space of the spinless Josephson effect: a) Rabi regime
$\omega_0/\epsilon N=1.2$; b) Intermediate regime $\omega_0/\epsilon N=0.6$; c)
Josephson regime $\omega_0/\epsilon N=0.4$.

FIG. 2. $V_{eff}(\Delta N_+)$ for different initial conditions: a) when the coefficient
of the quadratic term in Eq. (\ref{eq:Veff}), namely $\frac{1}{2} \omega_0^2 - \epsilon
 H(0) $, is positive; b) when the coefficient is
negative and $E_{eff}>0$; c) when both the coefficient and $E_{eff}$ are negative. The
horizontal line corresponds to $E_{eff}$.

FIG. 3. Equations (\ref{eq:equilibrium1}),(\ref{eq:equilibrium2}) plotted on the ($\Delta
N_1$, $\Delta N_2$) plane. The intersection of the two curves is the graphical solution
for the equilibrium points for $\Delta \varphi_1=\Delta \varphi_2=0$ (a, b), $\Delta
\varphi_1=\pi$ and $\Delta \varphi_2=0$ (c, d) and $\Delta \varphi_1=\Delta
\varphi_2=\pi$ (e, f, g).

TABLE 1. Equilibrium points and oscillation modes around them in the
isotropic case.

\pagebreak

{
\footnotesize
\setlength{\oddsidemargin}{-3.0cm}
\begin{tabular}{|c|c|c|c|c|c|}
\hline
$\Delta \varphi_1,\Delta \varphi_2$ & $\Delta N_1, \Delta N_2$ &
Existence condition & Type of mode & Frequency & Stability condition
\\
\hline
& & & density &
$\sqrt{\omega_0^2+\omega_0
\epsilon N}$ & 
\\
\raisebox{3.0ex}[0pt]{0,0} & \raisebox{3.0ex}[0pt]{0,0} &
\raisebox{3.0ex}[0pt]{always exists} & spin & $\omega_0$ &
\raisebox{3.0ex}[0pt]{always stable}
\\
\hline
& & & mixed & $\sqrt{\omega_0^2+\omega_0
\epsilon (N_2-N_1)}$ & $N_1-N_2-\omega_0/\epsilon <0$
\\
&\raisebox{3.0ex}[0pt]{0,0} &
\raisebox{3.0ex}[0pt]{always exists} &
spin & $\omega_0$ & always stable
\\
\cline{2-6}
\raisebox{3.0ex}[0pt]{0,$\pi$} & & & mixed &
$\epsilon|\Delta N_+^0|$ &
\\
& \raisebox{3.0ex}[0pt]{$\neq0$} &
\raisebox{3.0ex}[0pt]{$N_1-N_2-\omega_0/\epsilon>0$}
& spin & $\sqrt{\omega_0^2+\epsilon^2(\Delta N_+^0)^2}$
&\raisebox{3.0ex}[0pt]{always stable}
\\
\hline
& & & density & $\sqrt{\omega_0^2-\omega_0 \epsilon N}$ & $N_1+N_2- 
\omega_0/\epsilon<0$
\\
& \raisebox{3.0ex}[0pt]{0,0}
& \raisebox{3.0ex}[0pt]{always exists}
& spin & $\omega_0$ & always stable
\\
\cline{2-6}
\raisebox{3.0ex}[0pt]{$\pi,\pi$} & & & density & $\epsilon
|\Delta N_+^0|$ &
\\
& \raisebox{3.0ex}[0pt]{$\neq0$}
& \raisebox{3.0ex}[0pt]{$N_1+N_2-\omega_0/\epsilon>0$}
& spin & $\sqrt{\omega_0^2 + \epsilon^2 (\Delta N_+^0)^2}$ &
\raisebox{3.0ex}[0pt]{always stable}
\\
\hline
\end{tabular}
}

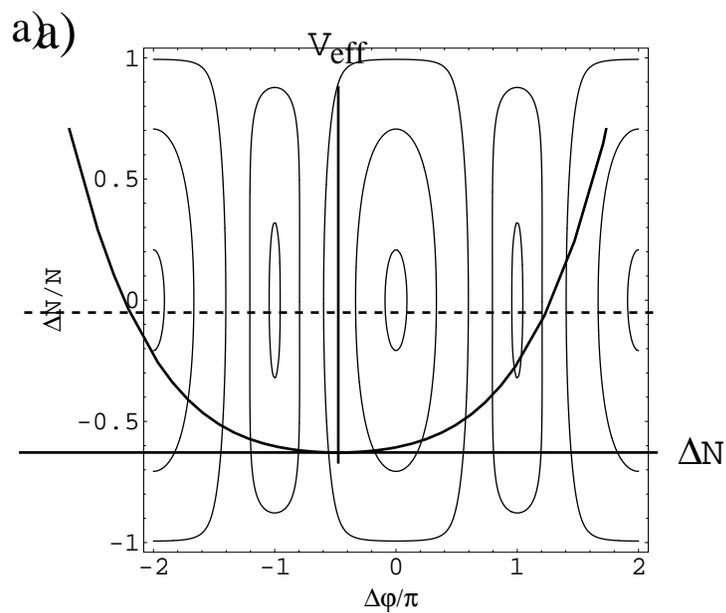
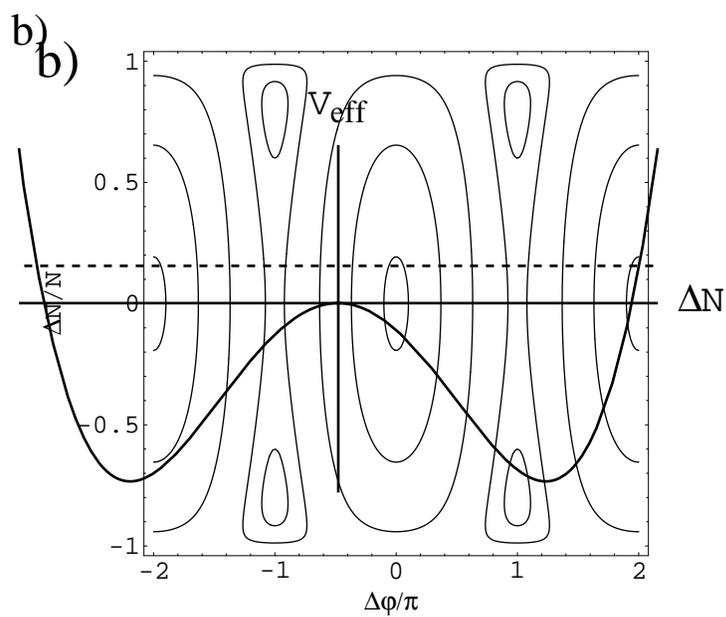
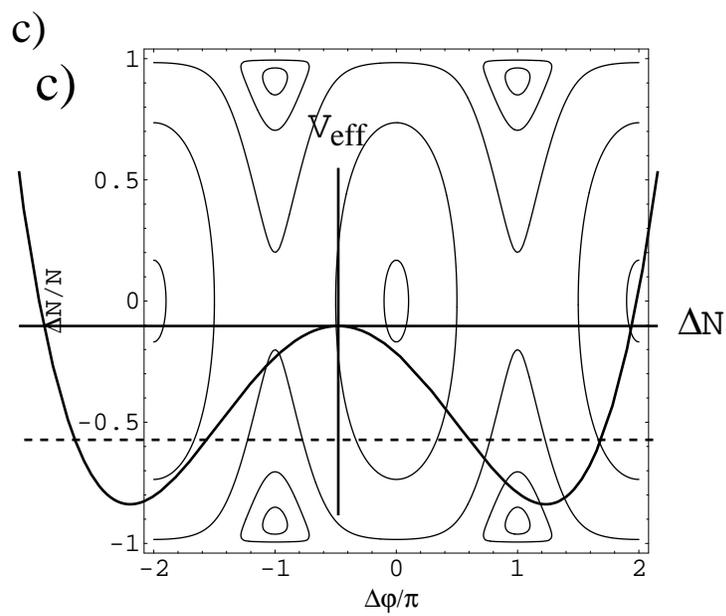
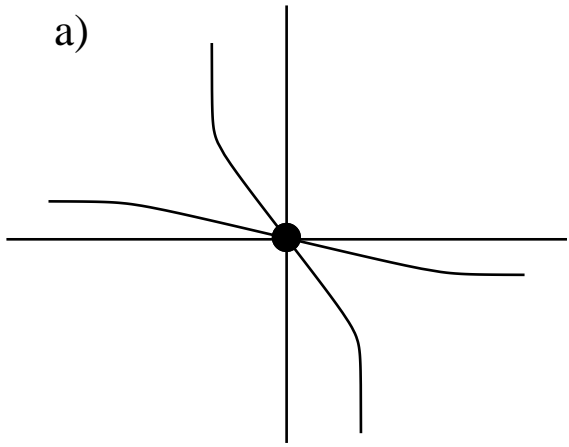
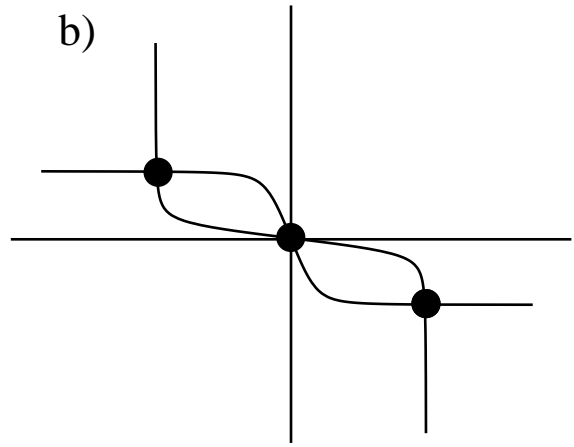
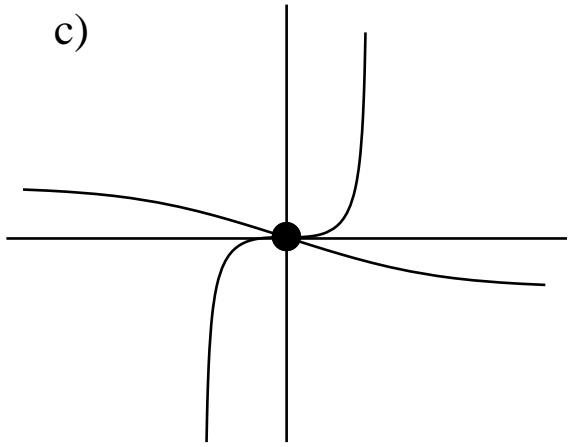
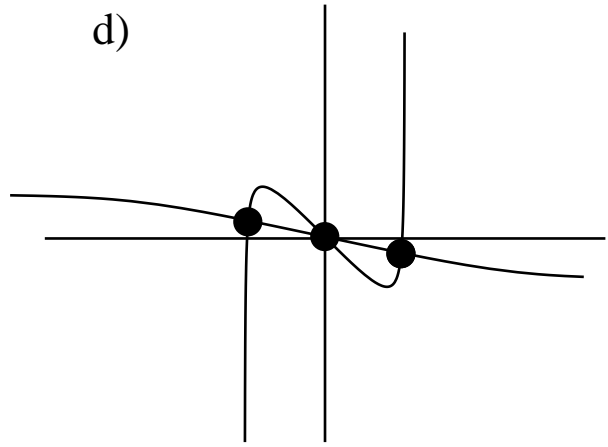
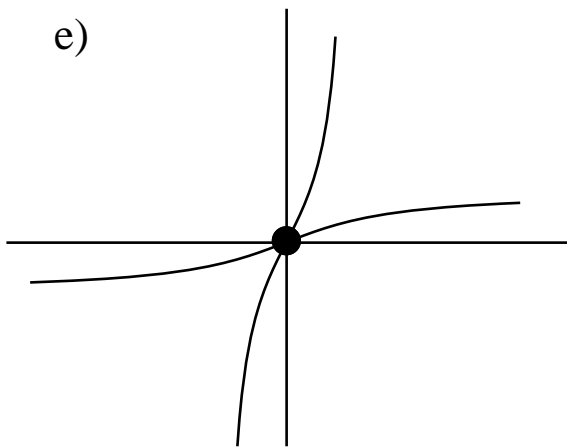
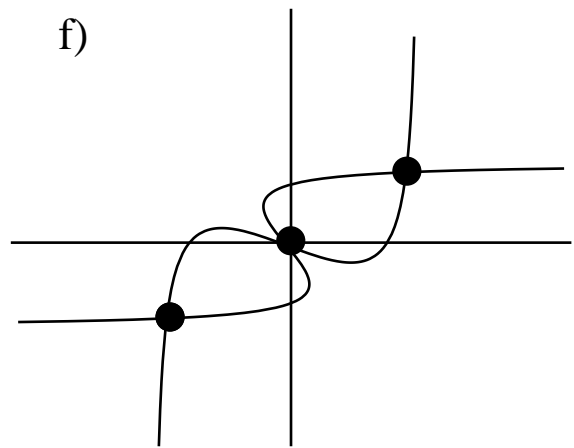
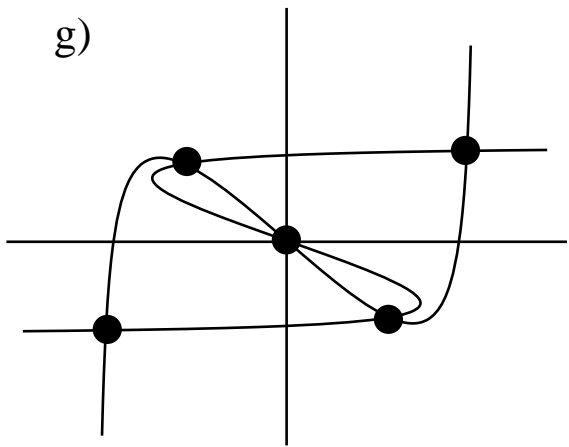
\end{document}